\documentclass[aps,prd,superscriptaddress,nofootinbib,amsmath,amsfonts,preprintnumbers,groupedaddress,showpacs,10pt,english]{revtex4-1}
\usepackage{amsmath}
\usepackage{amssymb}
\usepackage{babel}
\usepackage{wrapfig}
\usepackage{cancel}

\makeatletter

\usepackage{array,multirow,graphicx}
\usepackage{dcolumn}
\usepackage{newlfont}
\usepackage{bm}
\usepackage[colorlinks,citecolor=blue,urlcolor=blue,linkcolor=blue]{hyperref}
\usepackage[figtopcap]{subfigure}
\usepackage{color}

\begin{document}

\date{\today}

\preprint{FU-PCG-55}
\title{ Higher dimensional rotating  black {hole} solutions in quadratic $f(R)$ gravitational theory and the conserved quantities }
\author{G.G.L. Nashed$^{1}$}%
\email{nashed@bue.edu.eg}
\author{Kazuharu Bamba$^{2}$}%
\email{bamba@sss.fukushima-u.ac.jp}
\affiliation{$^{1}$Centre for Theoretical Physics, The British University in Egypt, P.O. Box 43, El Sherouk City, Cairo 11837, Egypt}
\affiliation{$^{2}$Division of Human Support System, Faculty of Symbiotic Systems Science, Fukushima University, Fukushima 960-1296, Japan}

\begin{abstract}
We explore the quadratic form of the $f(R)=R+bR^2$ gravitational theory to derive  rotating $N$-dimensions black   { hole} solutions with $a_i, i\geq 1$ rotation parameters. Here, $R$ is the Ricci scalar, and $b$ is the dimensional parameter. We assumed that the $N$-dimensional spacetime is static and has {flat  horizons} with a zero curvature boundary. We investigated the physics of black holes  by calculating the relations of  physical quantities such as the horizon radius and mass. We also demonstrate that in the four-dimensional case, the higher-order curvature does not contribute to the black {hole}, i.e.,  black hole { does} not depend on the dimensional parameter $b$  whereas in the case of $N>4$, it depends on  parameter $b$  owing to the contribution of the correction $R^2$ term. We analyze the conserved quantities, energy, and angular-momentum, of black {hole} solutions by applying the relocalization method. Additionally, we calculate the thermodynamic quantities such as temperature and  entropy  and examine the stability of  black {hole} solutions locally and show that they have thermodynamic stability. Moreover, the calculations of entropy  put a constraint on the parameter $b$ to be $b<\frac{1}{16\Lambda}$ to obtain a positive entropy.

\end{abstract}
\maketitle
\begin{center}
\section{\bf Introduction} \label{S1}
\end{center}

Einstein's general theory of relativity (GR) does not provide scientists an explanation  that can support the discovery of the  accelerated expansion of our universe which has been established 20-years ago   by  various observations \cite{Riess:1998cb,Perlmutter:1998np,Ade:2013zuv,Spergel:2006hy,Jain:2003tba,Cole:2005sx,Eisenstein:2005su,2010MNRAS.401.2148P,Padmanabhan:2012hf,2011MNRAS.418.1707B,2013MNRAS.428.1036M,PhysRevD.71.123001,Stern:2009ep,Zhang:2012mp,Blake:2011ep,Chuang:2012qt,Moresco:2012jh,2003MNRAS.346...78H,PhysRevD.69.103501,Cole:2005sx}. Thus, scientists have developed  other theories that can support this expansion rate. Among these theories, there is one in which we add a cosmological constant to the field equations of GR.  The output model of this theory becomes dominated by this constant which  can explain the accelerated expansion. In the literature this model is known as $\Lambda$ cold dark matter ($\Lambda$CDM). $\Lambda$CDM has a contradiction between gravity and quantum  field theory \cite{RevModPhys.61.1}. This leads scientists to modify  the structure of GR  either within Riemannian geometry \cite{Nashed:2018nll} or  using another geometry.

Among the modified theories that used other Riemannian geometry is the $f(T)$ theory, where $T$ is the torsion scalar in teleparallel gravity. This theory has second-order differential equations \cite{PhysRevD.75.084031,PhysRevD.78.124019,PhysRevD.79.124019}  makes it easy to analyze its  physics. $f(T)$ has  been used in the domain of cosmology \cite{PhysRevD.83.104017,Nashed:2015pda,PhysRevD.83.064035,Nashed:uja,Awad:2017sau} and in the solar system \cite{Nashed:2013bfa,Awad:2017tyz,Nashed:2016tbj,Capozziello:2012zj}. The other modified theories that used Riemannian geometry are as follows:\vspace{0.1cm}\\
i-String theory, which is one of the most possible candidates for the quantum theory for gravitation  \cite{Kallosh:2015nia}.\vspace{0.1cm}\\ ii-Lovelock theory,  which is the natural generalization of Einstein's GR to higher dimensions \cite{PhysRevD.91.064045}.\vspace{0.1cm}\\ iii- Brans-Dicke theory, whose  interaction is considered by GR tensor and scalar field  \cite{Kofinas:2016fcp}.\vspace{0.1cm}\\ v-$f(R)$ theory which we focus on in this study  \cite{PhysRevD.23.347,PhysRevD.77.023503} (for recent reviews on the dark energy problem and modified gravity theories to explain the late-time cosmic acceleration and inflation in the early universe, see, for instance,~\cite{Nojiri:2010wj, Capozziello:2011et, Capozziello:2010zz, Bamba:2015uma, Cai:2015emx, Nojiri:2017ncd, Bamba:2012cp}).

$f(R)$ theory has several successful applications in the domain of cosmology \cite{Artymowski:2014gea,PhysRevD.92.124024,Motohashi:2017aob,Huang:2013hsb,Addazi:2016bus}. However, it should be associated with other tests to achieve  the success of GR in the solar system \cite{Koyama:2015vza}. $f(R)$ is a modification of Einstein's GR and is considered as a novel geometrodynamical theory with degrees of freedom in the field equations of gravitation \cite{PhysRevD.79.124019,Utiyama:1962sn,1989PhLB..232..172B,Clifton:2006kc,PhysRevD.77.103523}. The action integral of this theory contains an appropriate function of the Ricci scalar $R$, and the field equations are of the fourth-order. The lower order of the field equations provides field equations of Einstein's GR, which are of the second order. A coincidence between $f(R)$ and other modified gravitational theories  through different frames can be found in \cite{Sk.:2017fac}.

{A viable inflationary model in gravity that considered  quantum corrections and included $R+R^2$ gravity as a particular case  was derived in \cite{Starobinsky:1980te}. The true forms of scalar and tensor perturbations created through inflation in such a model are discussed in \cite{1983SvAL....9..302S}.} The $f(R)$ theory can describe the inflationary stage and the dark energy-dominated stage, in which the late-time cosmic acceleration is realized \cite{2010RvMP...82..451S,DeFelice:2010aj,PhysRevD.90.024017}. A static spherically symmetric solution is discussed in \cite{Clifton:2006ug,Sebastiani:2010kv,Nashed:2019tuk} whereas studies on gravitational collapse can be found in \cite{Chakrabarti:2016aea,PhysRevD.94.104013}. Several black holes are derived in the framework of $f(R)$ \cite{PhysRevD.80.124011,Nashed:2018piz,Moon:2011hq,2018EPJP..133...18N,PhysRevD.94.024062,2018IJMPD..2750074N,Ca_ate_2016,Moon:2011fw,AyonBeato:2010tm,Hendi:2011eg,
 Hendi:2014mba,2013PhRvD..87j4029C} and their physical consequences are discussed in \cite{Addazi:2016hip,PhysRevD.91.064009,Akbar:2006mq,Faraoni:2010yi}.

The characteristics of gravitational theories in more than four dimensions have been studied  more widely \cite{Ortaggio:2004kr}.
The significant motivation of these studies is to identify the relation between black holes and fundamental theories,  such as string theory, besides to the consideration of the large extra dimensions with models of the TeV-scale gravity. The particular higher-dimensional solutions in classical GR have been found in the extensions to any $n > 4$ of the Schwarzschild and Reissner-Nordstr\"om black holes derived by Tangherlini \cite{Tangherlini1963} and that of the Kerr black hole solution analyzed by Myers and Perry \cite{1986AnPhy.172..304M}.
 However, new investigations have indicated that even at the classical level, higher-dimensional gravity theories have much richer dynamics for $n >4$. One of the most important features is the non-uniqueness of asymptotically-flat rotating black holes. For instance, for GR in a five-dimensional vacuum,
$S^1\times S^2$ rotating black ring solutions have been acquired explicitly \cite{Emparan:2001wn}. This object can have the same mass and spin as those of the $S^3$ black holes suggested in Ref.~\cite{1986AnPhy.172..304M}.
Such  violation will be infinite continuously for the rings with magnetic dipole charge \cite{Emparan:2004wy}.

To provide a good probe of the $f(R)$ gravitational theory, one has to analyze  black hole solutions. Exact solutions for $f(R)$ are a hot topic, and there are several studies on the topic starting from 3-dimensions \cite{Horne:1991gn} to $N$-dimensions \cite{Cisterna:2017qrb,2013PhyS...87d5004S}. Analytic solutions that describe rotating black holes are derived in \cite{PhysRevD.86.024013, Sheykhi:2013yga,PhysRevD.88.044044}.  The main purpose of this work is to derive solutions for  the $N$-dimension black {hole} with flat or cylindrical horizons in the framework of    $f(R)=R+bR^2$. To achieve  this, we  used a general N-dimension metric that possesses a $k-dimension$ Euclidean metric and derived a static black hole solution in diverse dimensions. Using a coordinate transformation, we successfully derive a stationary rotating black hole solution for $f(R)=R+bR^2$. The physics of these   black holes is  investigated by calculating  conserved quantities and studying thermodynamic quantities such as Hawking temperature, entropy, and heat capacity. The influence of higher-order curvature corrections on the existence of relativistic compact objects in modified gravity has been argued \cite{PhysRevD.78.064019,PhysRevD.79.024009}.

 This  paper is organized as follows: In  Section \ref{S2}, we provide the basics of the $f(R)$ gravitational theory and N-dimension  spacetime with one unknown function is  applied those to the quadratic form of $f(R)$ field equations. Additionally, in Section \ref{S2}, black {hole} solutions are  derived for two different cases, i.e. 4-dimensional case and $N>4$. In Section \ref{S3}, we apply a coordinate transformation to the black {hole} solutions derived in Section \ref{S2} and obtain rotating  black {hole} solutions that satisfy the field equations of the quadratic form of the  $f(R)$ theory. In Section \ref{S33}, we calculate the conserved quantities of the rotating solutions  using the Komar method and obtain divergent quantities.    In Section \ref{S4}, we use the regularized method and redo the calculations of the conservation and  obtain finite quantities for  rotating black holes.   Finally, in Section \ref{S5}, we discuss  the  stability of black {hole} solutions  locally  and   explain that the  derived solutions are stable from the viewpoint of thermodynamics.  In  Section \ref{S6}, we present our conclusions and discussion.\vspace{0.5cm}\\

\section{Basics of $f(R)$ gravitational theory} \label{S2}
We consider a gravitational field with a cosmological constant.  The action of this field is given by  \cite{Nashed:2018efg,Nashed:2019tuk,Nashed:2019yto,Elizalde:2020icc,Nashed:2020kdb}
\begin{equation}\label{m444}  S:=\frac{1}{2\chi} \int d^Nx \sqrt{-g} (f(R)-2\Lambda),\end{equation}
where $\Lambda$ is the cosmological constant and $\chi$ is the $N$-dimension gravity constant represented by $\chi =2(N-3)\Omega_{N-1} G_N$, where  $G_N$  is the  gravitation constant of Newton in $N$-dimensions. In this study, $\Omega_{N-1}$ shows the volume for the $(N-1)$-dimensional unit sphere. It is given by \cite{Nashed:2018efg,Nashed:2019tuk,Nashed:2019yto,Elizalde:2020icc}
\begin{equation} \Omega_{N-1} = \frac{2\pi^{(N-1)/2}}{\Gamma((N-1)/2)},\end{equation} where $\Gamma$ is the gamma function.

By varying  Eq. (\ref{m444})  with respect to the metric
tensor $g_{\mu \nu}$, the field equations for $f(R)$ can be derived in the following form  \cite{Cognola:2005de,Koivisto:2005yc}:
\begin{equation} \label{fe}
E_{\mu \nu}\equiv R_{\mu \nu} f_R-\frac{1}{2}g_{\mu \nu}f(R)-\frac{1}{2}g_{\mu \nu}\Lambda +g_{\mu \nu} \Box f_R-\nabla_\mu \nabla_\nu f_R-2\kappa T_{\mu \nu}=0,\end{equation}
where $R_{\mu \nu}$ is the Ricci tensor given by
 \begin{equation} \label{rt} R_{\mu \nu}=R^{\rho}{}_{\mu \rho \nu}=\partial_\rho\Gamma^\rho{}_{\mu \nu}-\partial_\nu\Gamma^\rho{}_{\mu \rho}+\Gamma^\rho{}_{\rho \beta}\Gamma^\beta{}_{\mu \nu}-\Gamma^\rho{}_{\nu \beta}\Gamma^\beta{}_{\mu \rho}= 2\Gamma^\rho{}_{\mu [\nu,\rho]}+2\Gamma^\rho{}_{\beta [\rho}\Gamma^\beta{}_{\nu] \mu},\end{equation}
 and $\Gamma^\rho{}_{\mu \nu}$ is the Christoffel symbols of the second kind and the square brackets mean skew-symmetrization.  The D'Alembert operator $\Box$ is defined as $\Box= \nabla_\alpha\nabla^\alpha $, where $\nabla_\alpha V^\beta$ is the covariant derivatives in terms of the vector $V^\beta$, $f_R=\frac{df(R)}{dR}$, and $T_{\mu \nu}$ is the energy-momentum tensor. The trace of  field  equations (\ref{fe}), in the vacuum case, gives
\begin{equation} \label{fe1}
Rf_R-\frac{N}{2}f(R)-8\Lambda+3\Box f_R=0.
\end{equation}
Equation (\ref{fe1}) with $f(R)=R$ gives $R=-8\Lambda$.

We will apply the field equations (\ref{fe}) to the following metric:
\begin{equation} \label{m2}
ds^2= -h(r)dt^2+\frac{1}{h(r)}dr^2+r^2\left(\sum_{i=1}^{\ell}d\phi^2_i+\sum_{k=1}^{N-\ell-2}dz_k{}^2\right).
\end{equation}
Here, $0\leq r< \infty$, $-\infty < t < \infty$, $0\leq \phi_{\ell}< 2\pi$,
$-\infty < z_k < \infty$, and $h(r)$\footnote{{ In this study, we consider spacetime (\ref{m2}) with one unknown function only to make the calculations more applicable. This constraint makes the spacetime not affected by the parameter $b$ in the 4-dimension because of the non-contribution of the $R^2$ term. To let the spacetime be affected by the parameter $b$, in the 4-dimensional case, we must consider the charged  form of the  field equation (\ref{fe}) \cite{2018IJMPD..2750074N}.}} is an unknown function in terms of the radial coordinate $r$.
 Using Eq. (\ref{m2})  we obtain the Ricci scalar as
\begin{equation}
R=-\frac{r^2 h''+2(N-2)r h'+(N-2)(N-3)h}{r^2},\end{equation} where $h'=\frac{dh(r)}{dr}$ and $h''=\frac{d^2h(r)}{dr^2}$.  The non-zero components of the $f(R)$  field equations, Eqs. (\ref{fe}), for $f(R)=R+b R^2$, where $b$ is a dimension parameter, and $T_\mu{}^\nu=0$, take the form\footnote{The detailed calculations of the Ricci curvature tensor are given in Appendix B.}
 \begin{eqnarray} \label{fe5}
&&E_t{}^t= \frac{1}{2r^4}\Bigg(b r^3\Bigg\{2h'''[rh'+6h(N-2)]+4rh h''''-rh''^2\Bigg\}+2(N-2)r^2b h''(2[3N-11]h+rh')+2(N^2-7N+10)b r^2 h'^2\nonumber\\
&&-h'[(N-2)r^3-2b r h(N-2)(3N^2-29N+64)]-h(N^2-5N+6)r^2+2b(N^2-N-2)h^2+4 r^4\Lambda\Bigg)=0,\nonumber\\
&& E_r{}^r= \frac{1}{2r^4}\Bigg(b r^3\Bigg\{2h'''[rh'+2h(N-2)]-rh''^2\Bigg\}+2(N-2)r^2b h''(4[N-2]h+rh')+2(N^2-7N+10)b r^2 h'^2\nonumber\\
&&-h'[(N-2)r^3+2\{4(N-2)-3(N-4)(N^2-5N+6)\}b r h]-(N-2)(N-3)[hr^2-b\{N^2-13N+22\}h^2]+4 r^4\Lambda\Bigg)=0,\nonumber\\
&& E_{\phi_1{}^{\phi_1} }=E_{\phi_2{}^{\phi_2}} \cdots E_{\phi_{N-\ell-2}{}^{\phi_{N-\ell-2}}=E_{z_1{}^{z_1}}=E_{z_2{}^{z_2}}=\cdots E_{z_{N-2}{}^{z_{N-2}}}} =\frac{1}{2r^4}\Bigg(b r^3\Bigg\{4h'''[rh'+h(3N-7)]+4rh h''''+rh''^2\Bigg\}\nonumber\\
&&-r^2 h''[r^2-2b \{2(3N-7)rh'-[2(N-2)-(N-4)(7N-15)]h\}]+4(N-2)[2N-9]b r^2 h'^2-(N-3)(N-4)r^2h+4 r^4\Lambda\nonumber\\
&&-h'(2(N-3)r^3+8(N-2)\{2(N-3)-(N-4)(N-5)\}b r h)-2b(N-6)[(N-3)(2N-1)-2(N-4)(N-5)]h^2\Bigg)=0.\nonumber\\
 &&
 \end{eqnarray}
The abovementioned system cannot have a general solution because of the appearance of the term (N-4). Therefore, we deal with it in  two separate cases.
The first case is the 4-dimension one in which
the solution for the abovementioned system is expressed as
\begin{eqnarray} \label{4d}
  h(r)=\frac{2r^2\Lambda}{3}+\frac{c_1}{r}.
 \end{eqnarray} Equation (\ref{4d}) shows that higher curvature has no effect, i.e., solution (\ref{4d}) is identical to GR.
  The second case  is the one in which {$N>4$},  and its solution takes the form
 \begin{eqnarray}  \label{nd}
 &&  h(r)=\frac{(N-2)r^2\left[1+\sqrt{1-\frac{16N(N-4)b \Lambda}{(N-2)^2}}\right]}{2N(N-1)(N-4)b}+\frac{c_2}{r^{N-3}}\equiv r^2 \Lambda_{eff}+\frac{c_2}{r^{N-3}},\nonumber\\
 && {\textrm where} \quad \Lambda_{eff}=\frac{(N-2)\left[1+\sqrt{1-\frac{16N(N-4)b \Lambda}{(N-2)^2}}\right]}{2N(N-1)(N-4)b}.
 \end{eqnarray}
where  $c_1$ and $c_2$ are the integration constants. {  Equation (\ref{nd}) shows how the solution is affected by the dimension parameter $b$. Also,  Eq. (\ref{nd}) informs us that the parameter $b$ should  not be equal to 0. In the case where we set the explicit cosmological constant $\Lambda=0$ in Eq. (\ref{nd}), we obtain
\begin{eqnarray}  \label{nd7} \Lambda_{eff}=\frac{(N-2)}{2N(N-1)(N-4)b}.\end{eqnarray} which shows that the parameter $b$  is an effective cosmological constant \cite{Nashed:2020kdb,Nashed:2018cth}. Hence, the metric potential (\ref{nd}) is new  and cannot be reduced to the GR metric when the dimensional parameter $b=0$.}
  The metric spacetimes of solutions (\ref{4d}) and (\ref{nd}) have the form
\begin{eqnarray} \label{m5}
ds_1{}^2&=&-\left\{\frac{2r^2\Lambda}{3}+\frac{c_1}{r}\right\}dt^2+\left\{\frac{2r^2\Lambda}{3}+\frac{c_1}{r}\right\}^{-1}dr^2+r^2\left(d\phi^2_1+dz^2_1\right)\;, \qquad \qquad \qquad N=4,\nonumber\\
 {ds_2{}^2}&=&{-\left\{r^2 \Lambda_{eff}+\frac{c_2}{r^{N-3}}\right\}dt^2+\left\{r^2 \Lambda_{eff}+\frac{c_2}{r^{N-3}}\right\}^{-1}dr^2+r^2\left(\sum_{i=2}^{n}d\phi^2_i+\sum_{k=2}^{N-n-2}dz_kdz_k\right)}\;,    \qquad  N>4. \end{eqnarray}
  {We stress the fact that solution  (\ref{nd})  is  new  for $N>4$ because it contains the dimensional parameter $b$, and cannot reduce to GR because  $b$, in this case, is not allowed to take zero value.}

  The asymptote of Eq. (\ref{m5}) behaves  as AdS/dS. We must stress the fact that the disappearance of the dimensional parameter $b$ in the 4-dimensional case is because we deal with a static black {hole}, however, if we study a charged black {hole}, this parameter should appear in the 4-dimensional case. Moreover, if we change the 4-dimensional case to the Einstein frame, the black {hole}, in that case will depend on the parameter $b$ because  the conformal factor $\sigma^2=f_{R}=1-16b\Lambda$ \cite{DeFelice:2010aj,Bahamonde:2017kbs,Bahamonde:2016wmz} and the metric will change owing to this transformation as \cite{DeFelice:2010aj} \begin{eqnarray} \label{conf-trans}
g_{\mu \nu} \to  {\bar g}_{{\mu \nu}_{Ein}}(x)=\sigma^2(x) g_{{\mu \nu}_{Jor}}(x),
\end{eqnarray}
where $ {\bar g}_{{\mu \nu}_{Ein}}(x)$ is the metric in Einstein frame while $ {\bar g}_{{\mu \nu}_{Jor}}$ is the one in Jordan frame.

\section{Rotating black {hole} solutions}\label{S3}
In this section, we analyze the rotating solutions, which satisfy the quadratic form of the field equations (\ref{fe}). To execute it, we explore two cases separately: \vspace{0.1cm}\\
i-The rotating case when $N=4$, \hspace{3cm} ii-The rotating case when $N>4$.\vspace{0.3cm}\\
\underline{i-The rotating case when $N=4$  }:\vspace{0.2cm}\\
In this case, we apply the following coordinate transformations\footnote{From now on, in the case of $N=4$, we write the cosmological constant in  the form of $\Lambda=-\frac{3}{\lambda^2}.$}
\begin{equation} \label{t3}
{\phi}'_1 =\frac{ a_1}{\lambda^2}~t-\Xi~ {\phi_1},\qquad \qquad \qquad {t}'= \Xi~ t-a_1~ \phi_1, \end{equation}  where  $a_1$ is the rotation parameter and  $\Xi$ is defined as
\[\Xi:=\sqrt{1+\frac{a_1{}^2}{\lambda^2}}.\]
With the transformation (\ref{t3}) to the metric (\ref{m5}) in the case of $N=4$, we acquire
\begin{eqnarray}  \label{m11}
&& ds_1{}^2=-\Big(\frac{\Xi^2 \lambda^2 h(r)-a_1{}^2 r^2}{\lambda^2}\Big)dt'^2 +\frac{dr^2}{h(r)}+r^2\left(\Xi^2d\phi'^2_1+dz^2_1\right)-a_1{}^2h(r)d\phi'^2_1+\frac{2\Xi a_1[r^2+\lambda h(r)]d\phi'_1 dt'}{\lambda}\;,\nonumber\\
&&  \end{eqnarray} where $h(r)$ is given by Eq. (\ref{4d}).
It should be mentioned that, for the rotation parameter $a_1=0$, we return to the spacetime (\ref{m5}) with $N=4$.

For $N>4$, we apply the following coordinate transformations\footnote{From now on, in the case of $N>4$,  we will write the cosmological constant, in the form of $\Lambda_{eff}=-\frac{(N-1)(N-2)}{(2\lambda_1{}^2)}.$}:
\begin{equation} \label{t1}
{\phi}'_i =-\Xi~ {\phi}_i+\frac{ a_i}{\lambda_1{}^2}~t,\qquad \qquad \qquad {t}'= \Xi~ t-\sum\limits_{i=2}^{{\ell}} a_i~ \phi_i, \end{equation} where $a_i$, $i>1$ is the number of rotation parameters and $\Xi$ is defined as
\[\Xi:=\sqrt{1+\sum\limits_{i=1}^{{\ell}}\frac{ a_i{}^2}{\lambda_1{}^2}}.\]
Applying the transformation (\ref{t1}) to the metric (\ref{m5}) in the case of $N>4$, we obtain
\begin{eqnarray} \label{m1}
&& ds_2{}^2=-h(r)\left[\Xi d{t'}  -\sum\limits_{i=2}^{\ell}  a_{i}d{\phi'} \right]^2+\frac{dr^2}{h(r)}+\frac{r^2}{\lambda_1{}^4}\sum\limits_{i=1 }^{\ell}\left[a_{i}d{t'}-\Xi \lambda_1{}^2 d{\phi'}_i\right]^2+ r^2dz_k{}^2\nonumber\\
&&-\frac{r^2}{\lambda_1{}^2}\sum\limits_{i<j }^{\ell}\left(a_{i}d{\phi'}_j-a_{j}d{\phi'}_i\right)^2\,,\end{eqnarray}  where $h(r)$ is given by Eq. (\ref{nd}),  $dz_k{}^2$ is
the Euclidean metric in (N-$\ell$-2)-dimensions and $k = 1,2\cdots N-2$.

 Of note, the static configuration (\ref{m5}) can be recovered as a special case when the rotation parameters $a_i$ equal to 0. It should  stressed that  when the physical quantities $c_1$ for $N=4$ and $c_2$ for $N>4$ are vanishing, we  obtain an odd  AdS spacetime\footnote{{Of note,  black hole solutions (\ref{m11}) and (\ref{m1}) are created by coordinate transformation  and are new because  they involve a rotation term which will be responsible for creating non-vanishing value of spatial momentum, as we will be discussing in Section \ref{S4}.} }.

Finally, it should be stressed that coordinate transformations (\ref{t3}) are admitted locally but not globally \cite{Lemos:1994xp, Awad:2002cz} because    the compactified angular coordinate $\phi$ is mixed with the temporal coordinate $t$ by  coordinate transformations.
This fact has been discussed in \cite{Stachel:1981fg} by pointing out that
if the first Betti number for the manifold is a non-zero value,
the global diffeomorphisms, by which two spacetimes can be connected,
do not exist thus there is a new manifold that is parameterized globally by the rotation parameters $a_i$.
The solution given by Eq. (\ref{m11}) shows that
the first Betti number is one of these solutions for the cylindrical or toroidal horizons. The same analysis can be applied to the coordinate transformation (\ref{t1}), for which the first Betti non-zero number is derived by the compactification of certain numbers of the angular coordinates in $(N-2)$-dimensional, $\phi_i$, to the submanifold of the solution.
In this study, we call these coordinates as rotation parameters of the solution.

\section{Total conserved charge}\label{S33}
We study the conserved quantities of the solutions found in the preceding section. For this purpose, we present the bases of the Einstein-Cartan geometry used for these calculations\footnote{Because the Ricci scalar of solutions (\ref{m11}) and (\ref{m1}) is equal to -$8\Lambda$ and $-8\Lambda_{eff}$, respectively, we are going to use the Komar formula to calculate the conserved quantities of the solutions derived in  Section  \ref{S3}.}. The Lagrangian of this theory has the form \cite{PhysRevD.74.064002}\footnote{The basic notations are given in Appendix A.}:\begin{equation} \label{lag}
{V(\vartheta^i, \
{R^j}_{k}):=-\frac{1}{2\kappa}\left(R^{i j}\wedge
\eta_{i j}-2\Lambda \eta\right)},\end{equation} where ${\vartheta^i}$ is the coframe, $\eta_{i j}$ is the basis of 2-form  and ${ R^{i j}}$ the Ricci that are one and two forms respectively.  Using the variational principle of Eq. (\ref{lag}), we obtain \cite{PhysRevD.74.064002,Kopczynski:1990af} \begin{eqnarray} && { E_{i}:= -\frac{\partial V}{\partial
\vartheta^i}=-\frac{1}{2\kappa}\left(R^{ j k}\wedge
\eta_{i j k}-2\Lambda \eta_i\right) ,   \qquad \qquad  B_{i j}:= -\frac{\partial
V}{R_{i j}}=\frac{1}{2\kappa}\lambda_{i j}},\end{eqnarray}
where ${B_{i j}}$ and $ { E_{i}}$ refer to the
 rotational gauge and energy-momentum, respectively. We can also define the following quantities
 \begin{eqnarray}  && E_{i j}:= -\vartheta_{[i}\wedge H_{j]}=0}, \qquad  { H_{i}:=-\frac{\partial
V}{\partial T^i}=0,\end{eqnarray}  which correspond to spin and translation, respectively. The conserved quantity is represented in the form \cite{PhysRevD.74.064002}
\begin{eqnarray} \label{con} && { \jmath[\xi]=\frac{1}{2\kappa}d\left\{^{^{*}}\left[dk+\xi\rfloor\left(\vartheta^i\wedge
T_i\right)\right]\right\}, \quad \textrm{where} \quad  k=\xi_i
\vartheta^i, \qquad \textrm{and} \qquad \xi^i=\xi\rfloor\vartheta^i}, \end{eqnarray}
where $\xi$ is a vector expressed as $\xi=\xi^i\partial_i$ with $\xi^i$
being $N$  parameters and $*$ is the Hodge duality. When the torsion one-form is vanishing, i.e., $T_i= 0$, the total charge of Eq. (\ref{con}) reads
\begin{equation} \label{con1} {{\cal Q}}[\xi]=\frac{1}{2\kappa}\int_{\partial S}{^*}dk. \end{equation}
This is the invariant conserved quantity  \cite{PhysRev.127.1411,Shirafuji:1996im,PhysRev.113.934,Ashtekar1980}.

We apply Eq. (\ref{con1}) to the solutions (\ref{m11}) and (\ref{m1}). We calculate the necessary components  for the case of $N=4$ and
 the co-frame is
\begin{eqnarray} \label{cof}
&& {{{\vartheta^{{}^{{}^{\!\!\!\!}}}}{_{}{_{}{_{}}}}}^{0}}=\sqrt{h(r)}[\Xi dt'-a_1d\phi'_1], \qquad {{{\vartheta^{{}^{{}^{\!\!\!\!}}}}{_{}{_{}{_{}}}}}^{\ 1}}=
 \frac{dr}{\sqrt{h(r)}}, \qquad {{{\vartheta^{{}^{{}^{\!\!\!\!}}}}{_{}{_{}{_{}}}}}^{\ 2}}=
 r dz_1, \qquad {{{\vartheta^{{}^{{}^{\!\!\!\!}}}}{_{}{_{}{_{}}}}}^{\ 3}}=r\Xi d\phi'_1-\frac{r a_1}{\lambda}dt'.\nonumber\\
 &&
\end{eqnarray}
With Eq. (\ref{cof}) and  Eq. (\ref{con}), we obtain \begin{equation} \label{kfor} k=\frac{\lambda^4 h^2(r)(a_1\xi_3-\Xi\xi_0)(\Xi dt'-a_1d\phi'_1) +\lambda^4 \xi_1dr+r^2h(r)[\lambda^4 \xi_2dz_1 +(\lambda^4 \Xi^2 \xi_3-\lambda^2 a_1\Xi \xi_0)d\phi'_1+a_1( a_1\xi_0- \lambda^2 \Xi \xi_3)dt']}{\lambda^4 h(r)}.\end{equation} The total derivative of Eq. (\ref{kfor}) gives

\begin{eqnarray} \label{dfor}
 && dk= \frac{1}{\lambda^4h(r)}\Bigg[\Bigg\{h'(r)\{\lambda^4\Xi( a_1\xi_3-\Xi  \xi_0)h(r)+r^2a_1(\lambda^2\Xi \xi_3-a_1\xi_0)\}-ra_1(\lambda^2 \Xi \xi_3-a_1\xi_0)[rh'(r)+2h(r)]\Bigg\}(dr \wedge dt')\nonumber\\
 && +2\lambda^4rh(r)\xi_2(dr \wedge dz_1)-\lambda^2\Bigg\{ h'(r)\{\lambda^2a_1( a_1\xi_3-\Xi  \xi_0)h(r)+r^2\Xi(\lambda^2\Xi \xi_3-a_1\xi_0)\} -r\Xi(\lambda^2\Xi \xi_3-a_1\xi_0)[rh'(r)+2h(r)]\Bigg\}\nonumber\\
 &&\times(dr \wedge d\phi'_1)\Bigg].
\end{eqnarray}
Using Eq. (\ref{cof}), we get
\begin{eqnarray} \label{cof1}
 && dt'=\frac{1}{r}\left( \frac{{{{\vartheta^{{}^{{}^{\!\!\!\!}}}}{_{}{_{}{_{}}}}}^{0}}r\Xi}{\sqrt{h(r)}}+\frac{{{{\vartheta^{{}^{{}^{\!\!\!\!}}}}{_{}{_{}{_{}}}}}^{\ 3}} a_1}{\lambda^2}\right), \qquad \qquad  d\phi'_1=\frac{1}{r}\left( {{{\vartheta^{{}^{{}^{\!\!\!\!}}}}{_{}{_{}{_{}}}}}^{3}}\Xi+\frac{{{{\vartheta^{{}^{{}^{\!\!\!\!}}}}{_{}{_{}{_{}}}}}^{\ 0}} ra_1}{\lambda^2\sqrt{h(r)}}\right), \qquad dr=
 {{{\vartheta^{{}^{{}^{\!\!\!\!}}}}{_{}{_{}{_{}}}}}^{\ 1}}\sqrt{h(r)}, \qquad dz_1=\frac{{{{\vartheta^{{}^{{}^{\!\!\!\!}}}}{_{}{_{}{_{}}}}}^{\ 2}}}{
 r} .\nonumber\\
 &&
\end{eqnarray}
By combining Eq. (\ref{dfor}) with Eq. (\ref{cof1}) and  Eq. (\ref{con1}) and operating the Hodge-dual to $dk$, we acquire the following forms for the total conserved charge
\begin{equation} \label{4dcon} { {{\cal Q}}[\xi_t']=\frac{\Xi^2}{\lambda^2}(M-4r^3) , \qquad \qquad {{\cal Q}}[\xi_r]={{\cal Q}}[\xi_{z_k}]=0, \qquad {{\cal Q}}[\xi_{\phi'_1}]=\frac{a_1 \Xi}{\lambda^2}(M-4r^3)},\end{equation}
where $M=-c_1$.
Throughout the same calculations for the case of $N>4$, with Eq. (\ref{m1}), we find
 \begin{eqnarray} \label{cofn}
&& {{{\vartheta^{{}^{{}^{\!\!\!\!}}}}{_{}{_{}{_{}}}}}^{0}}=\sqrt{h(r)}[\Xi dt'-\sum\limits_{i=1 }^{\ell}a_id\phi'_i], \qquad {{{\vartheta^{{}^{{}^{\!\!\!\!}}}}{_{}{_{}{_{}}}}}^{\ 1}}=
 \frac{dr}{\sqrt{h(r)}}, \qquad {{{\vartheta^{{}^{{}^{\!\!\!\!}}}}{_{}{_{}{_{}}}}}^{\ z_1}}=
 r dz_1, \qquad {{{\vartheta^{{}^{{}^{\!\!\!\!}}}}{_{}{_{}{_{}}}}}^{\ z_2}}=
 r dz_2, \cdots \qquad {{{\vartheta^{{}^{{}^{\!\!\!\!}}}}{_{}{_{}{_{}}}}}^{\ z_{N-\ell-2}}}=
 r dz_{N-\ell-2},\nonumber\\
 && {{{\vartheta^{{}^{{}^{\!\!\!\!}}}}{_{}{_{}{_{}}}}}^{\ \phi'_i}}=r\Xi d\phi'_i-\frac{r a_i}{\lambda^2}dt',\nonumber\\
\end{eqnarray} where $i=2\cdots \cdots \ell$.
By substituting Eq. (\ref{cofn}) into Eq. (\ref{con1}), we acquire \begin{eqnarray}  \label{cofnn1} &&  k=\frac{1}{\lambda^4 h(r)}\Bigg[h^2(r)\lambda^4(\sum\limits_{i=1 }^{\ell}a_i\xi_{i+k+1}-\Xi\xi_0)(\Xi dt'-\sum\limits_{i=1 }^{\ell}a_id\phi'_i) +\lambda^4 \xi_1dr+r^2h(r)(\lambda^4 \sum\limits_{i=1 }^{k}\xi_{i+1}dz_i\nonumber\\
&& +\sum\limits_{i=1 }^{\ell}(\lambda^4 \Xi^2 \xi_{{i+k+1}}-\lambda^2 a_i\Xi \xi_0)d\phi'_i+\sum\limits_{i=1 }^{\ell}( a_i{}^2\xi_0- \lambda^2 \Xi a_i\xi_{i+k+1})dt')\Bigg].\end{eqnarray} The total derivative of Eq. (\ref{cofnn1}) yields
\begin{eqnarray} \label{cofnnn}
 && dk= \frac{1}{\lambda_1{}^4h(r)}\Bigg[\Bigg\{h'(r)\Bigg[\lambda_1{}^4\Xi( \sum\limits_{i=1 }^{\ell}a_i\xi_{i+k+1}-\Xi  \xi_0)h(r)+r^2(\lambda_1{}^2\Xi\sum\limits_{i=1 }^{\ell}a_i\xi_{i+k+1} -\left(\sum\limits_{i=1 }^{\ell}a_i\right)^2\xi_0)\Bigg]-r(\lambda_1{}^2\Xi\sum\limits_{i=1 }^{\ell}a_i\xi_{i+k+1} \nonumber\\
 &&-\left(\sum\limits_{i=1 }^{\ell}a_i\right)^2\xi_0)[rh'(r)+2h(r)]\Bigg\}(dr \wedge dt')
 +2\lambda_1{}^4rh(r)\sum\limits_{i=1 }^{k}\xi_{i+1}(dr \wedge dz_i)-\lambda_1{}^2\sum\limits_{i=1 }^{\ell}(dr \wedge d\phi'_i) \Bigg\{ h'(r)\{\lambda_1{}^2 a_i( \sum\limits_{j=1 }^{\ell}a_j\xi_{j+k+1}\nonumber\\
 && -\Xi  \xi_0)h(r)+r^2\Xi(\lambda_1{}^2\Xi \xi_{i+k+1}-a_i\xi_0)\} -r\Xi(\lambda_1{}^2\Xi \xi_{i+k+1}-a_i\xi_0)[rh'(r)+2h(r)]\Bigg\}\Bigg].
\end{eqnarray}
We calculate the inverse of Eq. (\ref{cofn}), as we have done in the 4-dimensional case. By combining the results and Eq. (\ref{cofnnn}) and using
the Hodge-dual, we find that the conservation of $N>4$  of Eq. (\ref{m1}) is represented as
\begin{equation} \label{Ncon} {{\cal Q}}[\xi_t]=\frac{\Omega_{D-1}h'(r)}{16 \pi \lambda_1{}^2}, \qquad \qquad {{\cal Q}}[\xi_r]={{\cal Q}}[\xi_{z_i}]=0, \qquad                         {{\cal Q}}[\xi_{\phi_i}]=\frac{a_i h'(r) \Omega_{D-1} }{16 \pi \lambda_1{}^2},\end{equation} where $h(r)$ in the case of 4-dimensions is given by Eq. (\ref{4d}) and  $h'(r)=\frac{dh(r)}{dr}$ and in the case $N>4$ $h(r)$ is given by (\ref{nd}).

Equations (\ref{4dcon}) and (\ref{Ncon}) show that the conserved quantities of  spacetimes (\ref{m11}) and (\ref{m1}) are divergent. Thus,  regularization is necessary for Eq. (\ref{con1}).

\section{Regularization with relocalization for the conserved charge}\label{S4}
It is observed that for the general coordinate and local Lorentz transformations, Eq. (\ref{lag}) is invariant.
however, in \cite{PhysRevD.74.064002},  it has been indicated that
there is  one more vagueness in terms of the definition for the conserved
quantities rather than the diffeomorphism and  local Lorentz freedom.
This occurs because the relocalization in terms of the momenta of the gravitational field can always be allowed by field equations.
Relocalization is generated from the change of the Lagrangian for
the gravitational field in terms of the total derivative,
described as \begin{equation} \label{lag2}  V'=V+d\Phi, \qquad
\Phi=\Phi({{{\vartheta^{{}^{{}^{\!\!\!\!}}}}{_{}{_{}{_{}}}}}^{\alpha}},
{\Gamma_\alpha}^\beta, T^\alpha, {R_\alpha}^\beta).\end{equation}
The second term in the right-hand-side of Eq. (\ref{lag2}) is responsible for
the change of the boundary part for the action only; thus the field equations are left unaltered \cite{PhysRevD.74.064002}.
For the relocalization method, the conserved charge is \cite{PhysRevD.74.064002}
\begin{equation}  \label{conr} {{\cal J}}[\xi]=-\frac{\lambda^2}{4\kappa }\int_{\partial S}
\eta_{\alpha \beta \mu \nu}\Xi^{\alpha \beta} W^{\mu \nu}. \end{equation}
Here, $W^{\mu \nu}$ is the Weyl 2-form, given by \begin{equation} W^{\alpha
\beta}=\frac{1}{2}{C_{\mu \nu}}^{\alpha
\beta}{\vartheta}^{\mu}\wedge {\vartheta}^{\nu},\end{equation} where ${C_{\mu
\nu}}^{\alpha \beta}={h_\mu}^i{h_\nu}^j {h^\alpha}_k
{h^\beta}_l{C_{i j}}^{k l}$ is the Weyl tensor, and $\Xi^{\alpha \beta}$
is represented as\footnote{In \cite{PhysRevD.74.064002,PhysRevD.76.124030,Nashed:2011fg,Obukhov:2007sh},  explanations for the way of deriving Eq. (\ref{conr}) are provided.}
\begin{equation} \Xi_{\alpha \beta}=\frac{1}{2}e_\beta\rfloor
e_\alpha \rfloor dk.\end{equation}
It is known that for the coordinate and
local Lorentz transformations, the conserved currents ${{\cal J}}[\xi]$ do not
change. The vector field $\xi$ on the manifold of the spacetime is associated
 with the currents ${{\cal J}}[\xi]$.
To analyze the conserved quantities in terms of spacetimes
(\ref{m11}) and (\ref{m1}), Eq. (\ref{conr}) is used.

In the case of $N=4$, with the metric spacetime (\ref{m11}),
we examine the components of Eq. (\ref{conr}).
The non-zero components in terms of $\Xi^{\alpha \beta}$ read\footnote{
In Appendix, the non-zero components of the Weyl tensor are described.}
\begin{equation} \label{s1} \Xi_{01} =-\frac{[\Xi\xi_0+a_1\xi_3][c_1\lambda^2-4r^3]}{2r^2\lambda^2},\qquad \qquad \qquad
 \Xi_{1 3} =-\frac{[a_1\xi_0-\Xi\xi_3\lambda^2]\sqrt{h(r)}}{\lambda^2},
 \end{equation}

Using Eqs. (\ref{conr}) and (\ref{s1}), we obtain \begin{equation}  \label{con4} \eta_{\alpha \beta \mu \nu}\Xi^{\alpha \beta}
W^{\mu \nu}\cong-\frac{4c_1([a_1{}^2+2\lambda^2\Xi^2]\xi_0+a_1 \Xi\lambda^2\xi_3)}{\lambda^4}+O\left(\frac{1}{r^3}\right).\end{equation}
The substitution of Eq. (\ref{con4}) into Eq. (\ref{conr}) leads to \begin{equation} \label{s2} {{\cal
J}}[\xi_t]=M[3\Xi^2-1],\qquad {{\cal
J}}[\xi_r]={{\cal
J}}[\xi_\theta]=0, \qquad {{\cal
J}}[\xi_{\phi_1}]=Ma_1\Xi,\end{equation}
which is consistent with the result in \cite{Awad:2000gg,PhysRevD.86.024013}.

Throughout the same procedure for the metric spacetime (\ref{m1}), we acquire the following non-zero components in terms of $\Xi^{\alpha \beta}$
\begin{eqnarray} &&  \Xi_{1 t} =\left[\sum\limits_{i=0 }^{\ell}a_{1+i}\xi_{n-k+i}-\Xi\xi_0\right]h'(r),\qquad  \Xi_{1 n-k+i} =-\frac{2[a_{1+i}\xi_0-\Xi\xi_{n-k+i}\lambda_1{}^2]\sqrt{h(r)}}{\lambda_1{}^2},
  \end{eqnarray}
where $h(r)$ is given by Eq. (\ref{nd}).
Using Eqs. (\ref{conr}), we have \begin{equation} \label{con5} \eta_{\alpha \beta \mu \nu}\Xi^{\alpha \beta}
W^{\mu \nu}\cong\frac{4c_1\left([\sum\limits_{i=0 }^{\ell}a_i{}^2+(n-2)\lambda_1{}^2\Xi^2]\xi_0+\sum\limits_{i=0 }^{\ell}a_i \Xi\lambda_1{}^2\xi_3\right)}{\lambda_1{}^2}+O\left(\frac{1}{r^6}\right).\end{equation}
By combining Eqs. (\ref{con5}) and (\ref{conr}), we obtain \begin{equation} \label{conrr} {{\cal
J}}[\xi_t]=M[(n-1)\Xi^2-1]\xi_0,\qquad {{\cal
J}}[\xi_r]={{\cal
J}}[\xi_\theta]=0, \qquad {{\cal
J}}[\xi_{\phi_{1+i}}]=Ma_{1+i}\Xi\xi_{n-\ell+i},\end{equation}
where $i=1,2 \cdots \ell-1$ and have put $c_2=-M$.
Equation (\ref{conrr}) is compatible with what derived in \cite{Awad:2000gg}.

\section{Thermodynamics for black holes}\label{S5}
\begin{figure}
\centering
\subfigure[~The 4-dimensional case]{\label{fig:1a}\includegraphics[scale=0.35]{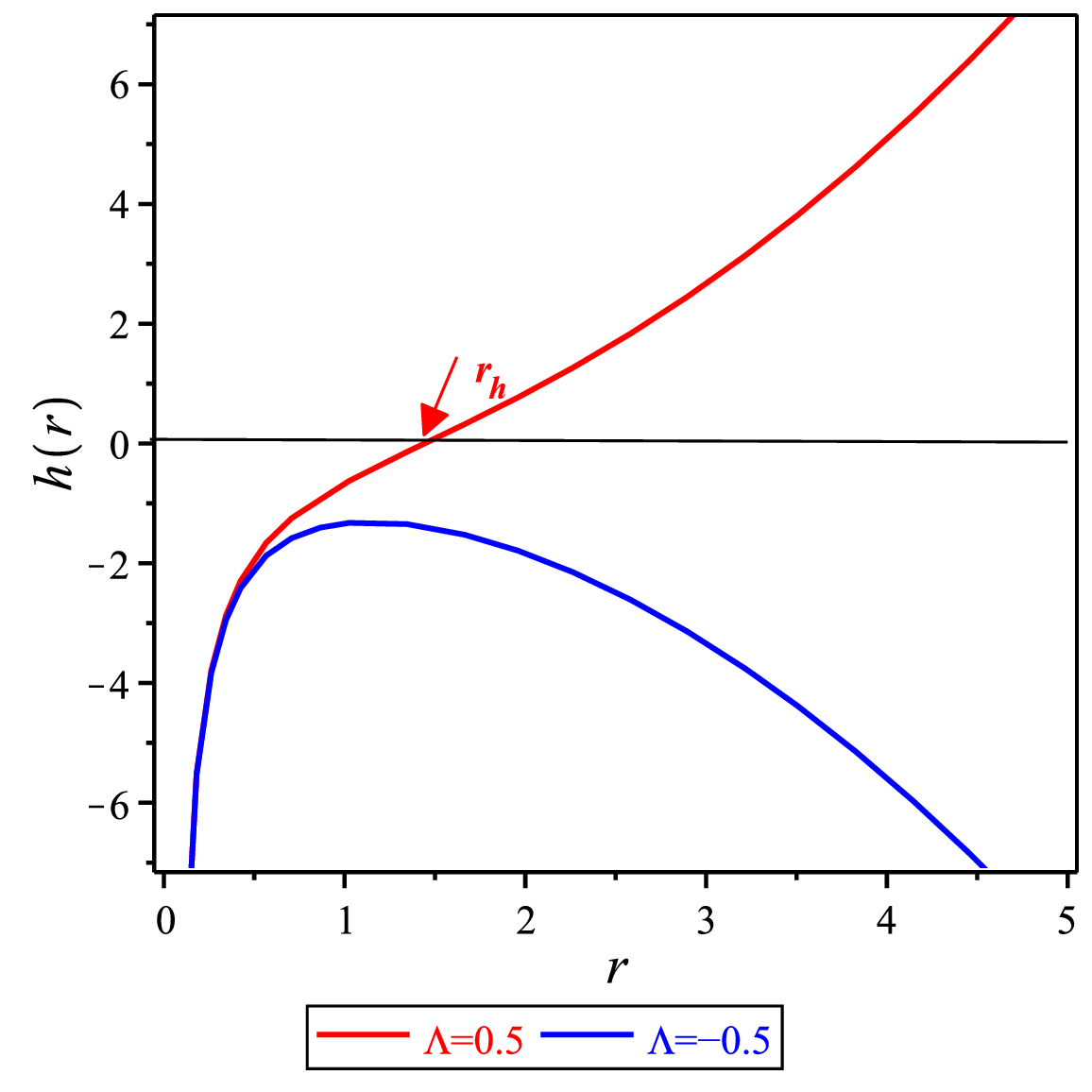}}\hspace{0.5cm}
\subfigure[~The 5-dimensional case]{\label{fig:1b}\includegraphics[scale=0.35]{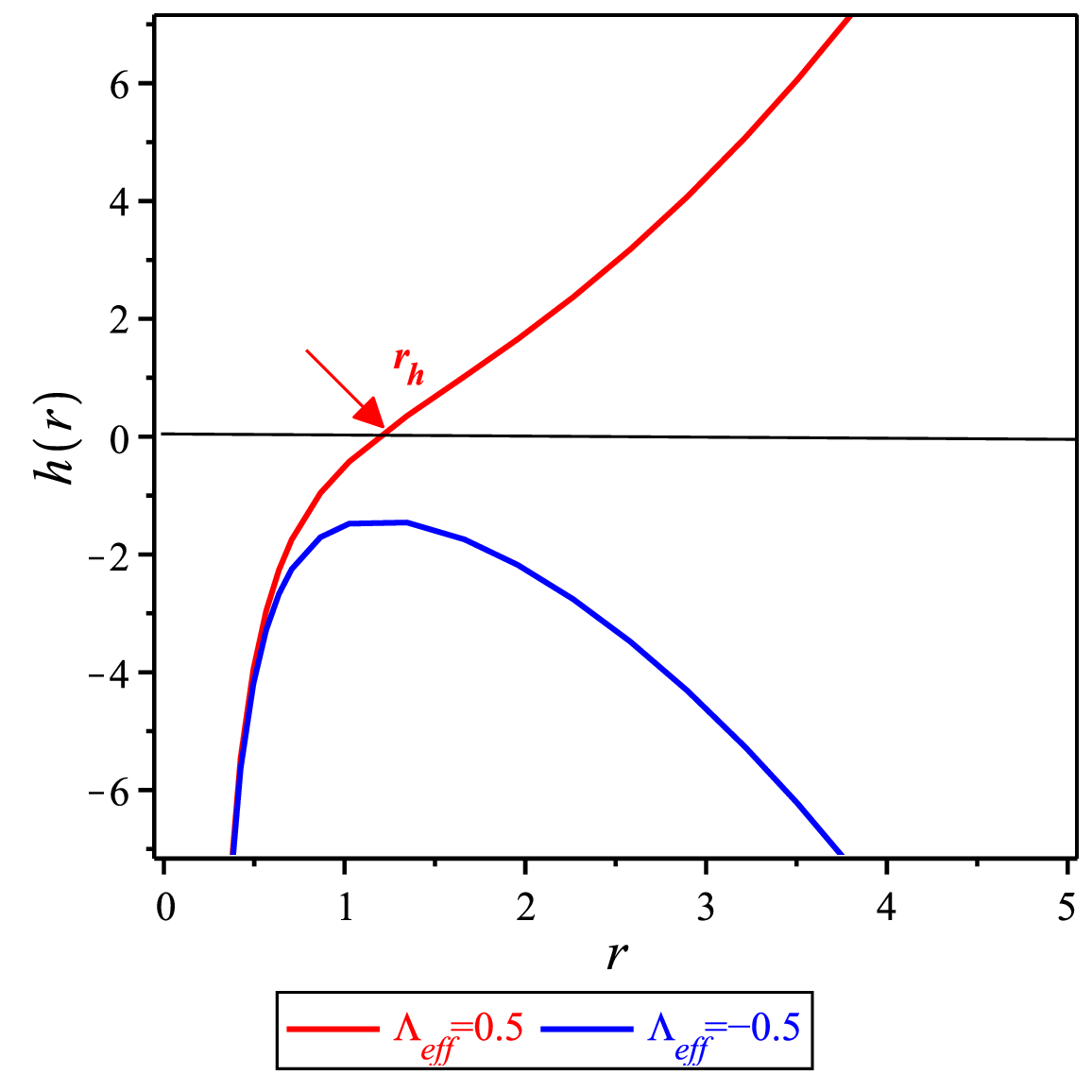}}
\caption{The function h(r) vs the radial coordinate $r$  for \subref{fig:1a} N=4,  $c_1 =-1$.\subref{fig:1b} N=5,  $c_2 =-1$\footnote{ {All figures are reproduced using the Maple software 16.}}. }
\label{Fig:1}
\end{figure}

In this section, we describe the  thermodynamic  quantities (e.g., temperature,  entropy and heat capacity)  of  the black hole solutions (\ref{4d}) and (\ref{nd}). For this, we define the Hawking temperature as \cite{PhysRevD.86.024013,Sheykhi:2010zz,Hendi:2010gq,PhysRevD.81.084040}:
  \begin{equation} \label{tem}
T_h = \frac{h'(r_h)}{4\pi}.
\end{equation}
Using Eqs. (\ref{4d}) and (\ref{nd}) in Eq. (\ref{tem}), we obtain
\begin{equation} \label{tem1}
{T_h=\frac{r_{h}\Lambda}{4\pi},\quad {\textrm when}\;  N=4,\qquad  T_h=\frac{{(N-1)r_{h} \Lambda_{eff} }}{4\pi}, \quad  {\textrm when}\; N>4},\end{equation} where $r_h$ is the largest root of the function $h(r)$ given by Eqs. (\ref{4d}) and (\ref{nd}), respectively. The relation between the function $h(r)$ and the radial coordinate $r$ for  black hole solutions (\ref{4d}) and (\ref{nd}) is plotted in Figures  \ref{Fig:1}\subref{fig:1a} and \ref{Fig:1}\subref{fig:1b},  which show that  we have an outer event horizon for the positive cosmological constant.

\begin{figure}
\centering
\subfigure[~The 4-dimensional case]{\label{fig:2a}\includegraphics[scale=0.25]{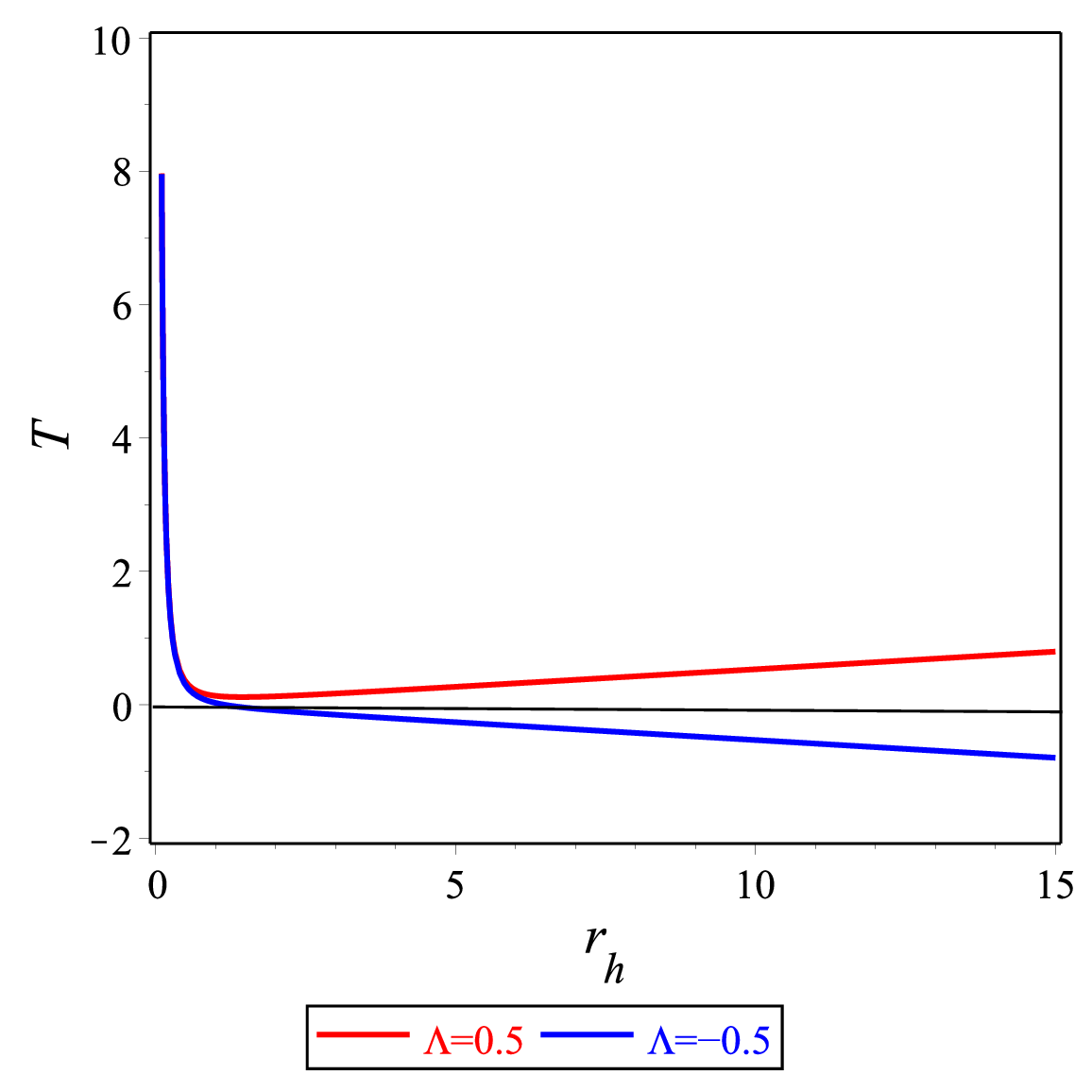}}\hspace{0.3cm}
\subfigure[~The 4-dimensional case]{\label{fig:2b}\includegraphics[scale=0.25]{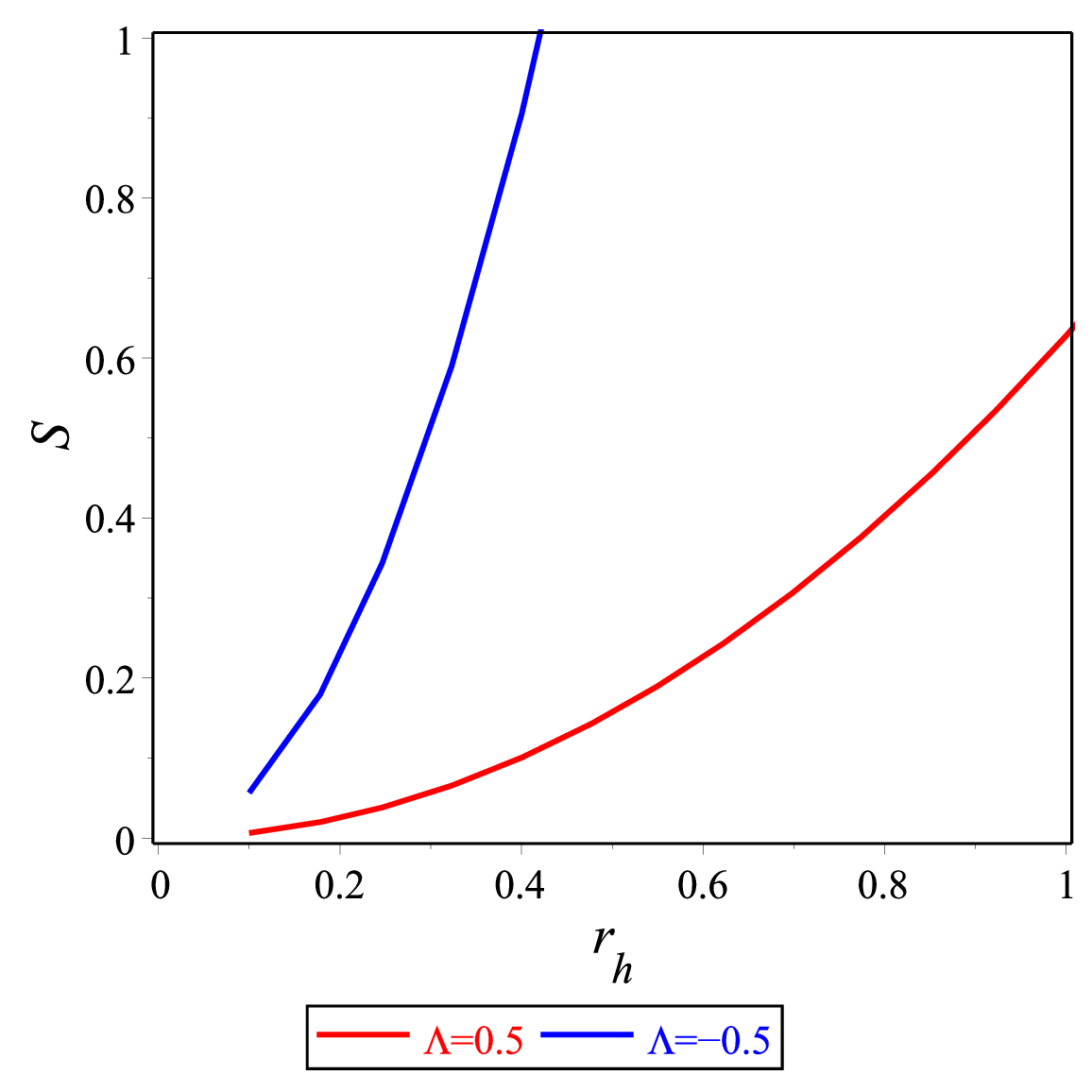}}\hspace{0.3cm}
\subfigure[~The 4-dimensional case]{\label{fig:2c}\includegraphics[scale=0.25]{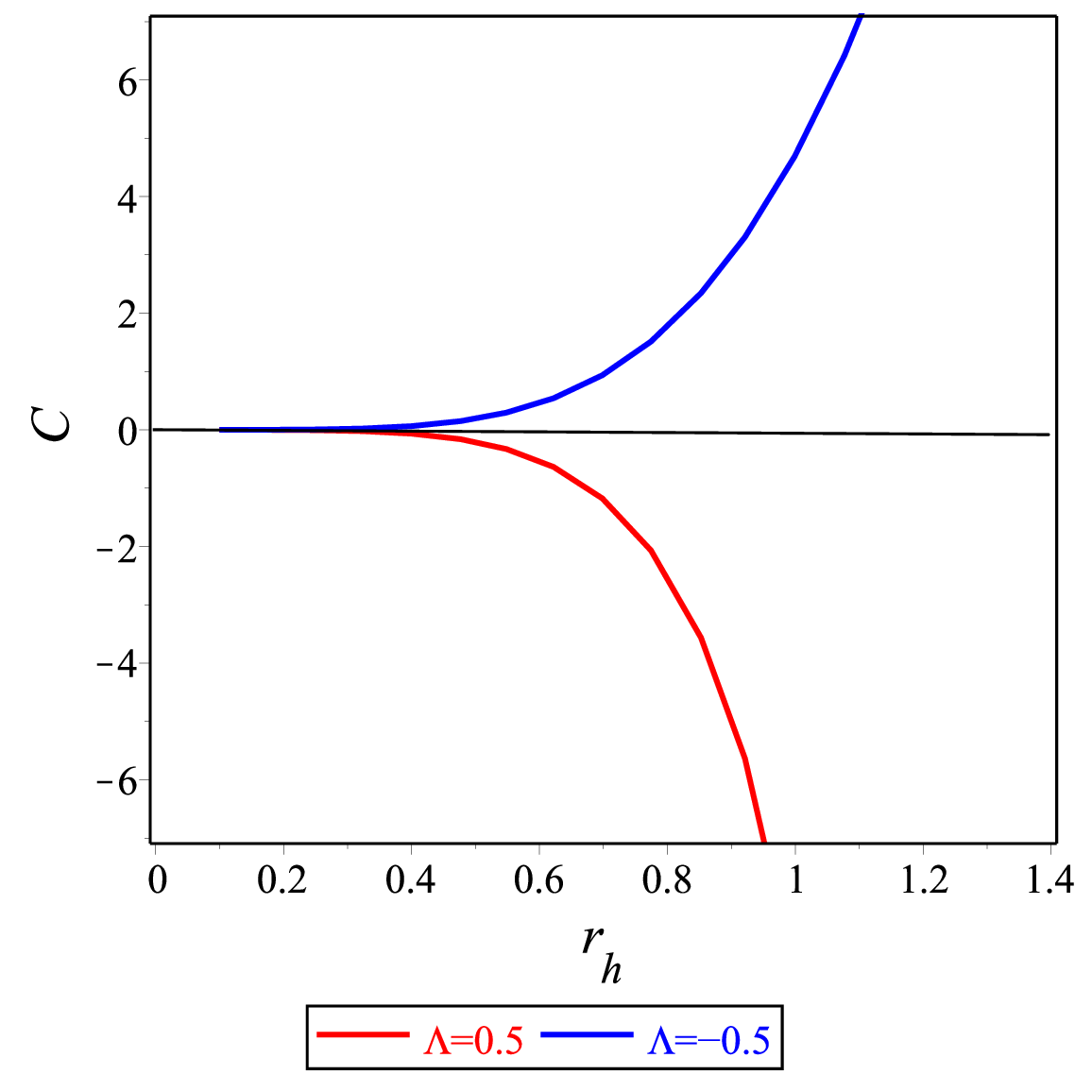}}\hspace{0.3cm}
\subfigure[~The 5-dimensional case]{\label{fig:2d}\includegraphics[scale=0.25]{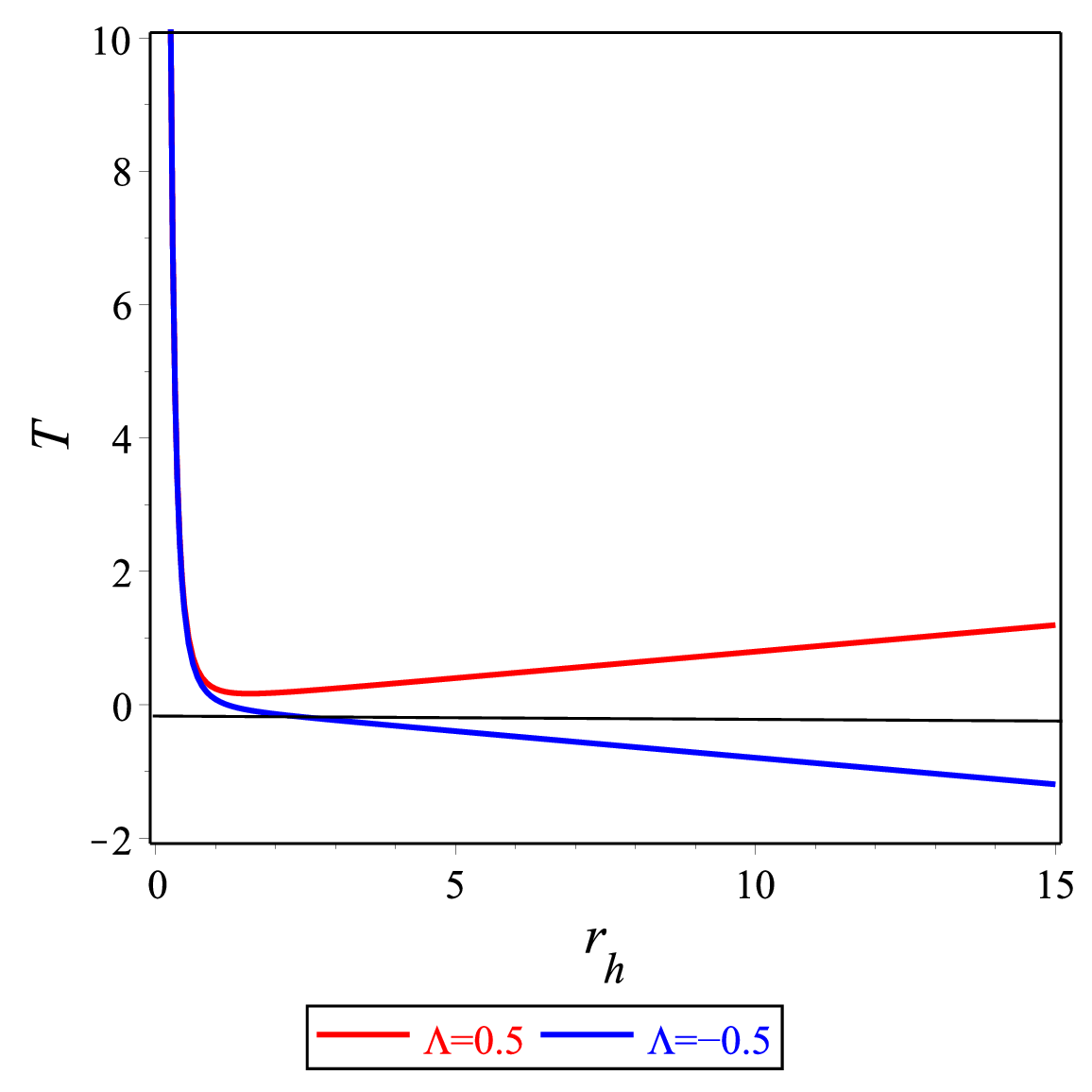}}\hspace{0.3cm}
\subfigure[~The 5-dimensional case]{\label{fig:2e}\includegraphics[scale=0.25]{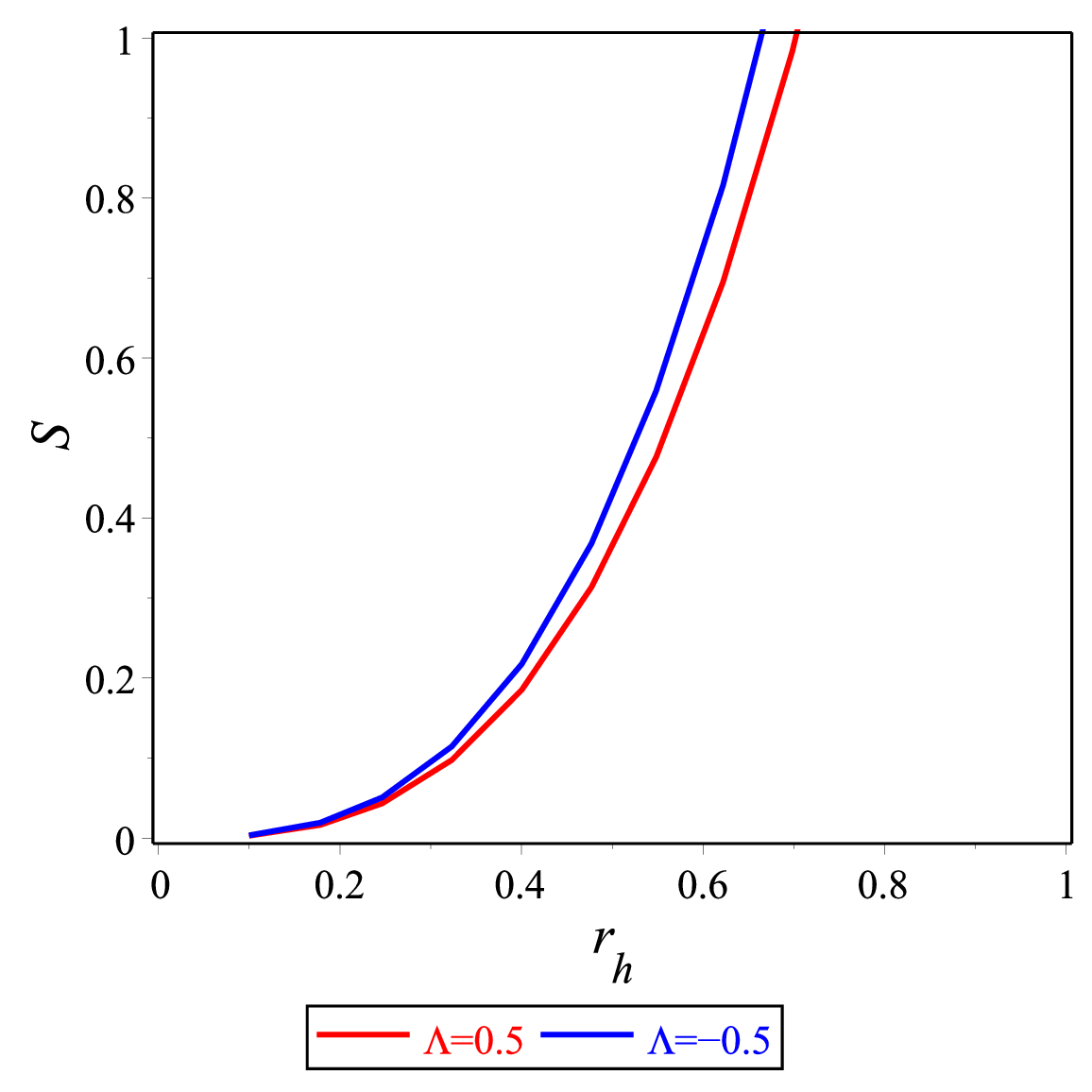}}\hspace{0.3cm}
\subfigure[~The 5-dimensional case]{\label{fig:2f}\includegraphics[scale=0.25]{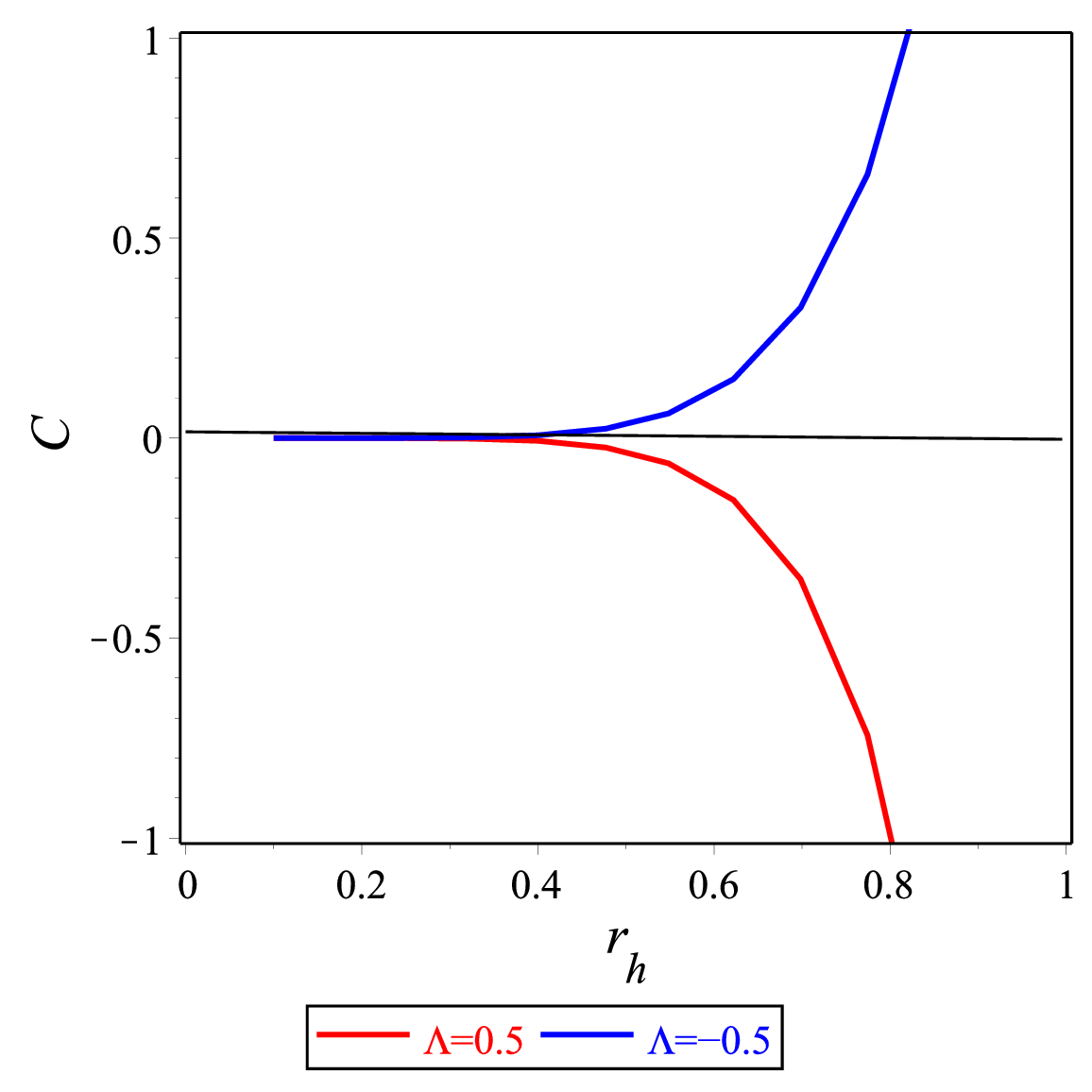}}
\caption{Horizon $r_h$ vs. \subref{fig:2a} and \subref{fig:2d} Hawking temperature \subref{fig:2b} and \subref{fig:2e}; entropy \subref{fig:2c} and \subref{fig:2f} heat capacity for the 4-dimensional and 5-dimensional cases,  respectively. In these figures, we take $b=0.1$ when $N=4$ and $b=0.01$ when $N=5$.  }
\label{Fig:2}
\end{figure}

{Also, in Figures \ref{Fig:2}\subref{fig:2a} and \subref{fig:2d}, we plot the behavior of  temperature vs. the horizon   for  $N=4$ and $N=5$. The abovementioned figures show that we have a positive temperature for positive $\Lambda$ and vice versa, i.e., a negative temperature for negative $\Lambda$.  This negative  Hawking temperature is responsible for forming an ultracold black hole. This result has been approved by Davies \cite{Davies:1978mf} who has shown that there is no logic in preventing a black hole temperature from having a negative value to switch it to a naked singularity. In fact, this is the case presented in Figures \ref{Fig:2} \subref{fig:2a} and \subref{fig:2d}. The case of an ultracold black hole can be explained by the existence of a phantom energy field \cite{Babichev:2014lda}. Moreover, it has been shown that the negative nature of
  temperature is related to  quantum properties \cite{Saridakis:2009uu}. }

To investigate the thermodynamic quantities in the context of  black hole solutions (\ref{4d}) and (\ref{nd}), we set the constraint $h(r_h) = 0$, which gives
\begin{eqnarray} \label{m33}
&& {M_h}_{{}_{{}_{{}_{{}_{\tiny Eq. (\ref{4d})}}}}}=\frac{2r_h{}^3 \Lambda}{3} \quad{\textrm for}\;  N=4,  \qquad {M_h}_{{}_{{}_{{}_{{}_{\tiny Eq. (\ref{nd})}}}}}=r_h{}^{N-1} \Lambda \quad{\textrm for}\;,  \quad {\textrm when}\; N>4 .
\end{eqnarray}
Now, we will briefly discuss the entropy
  of black hole in $f(R)$ gravity. For this, we use the arguments
presented in \cite{PhysRevD.70.043520}. From the Noether method which was used to calculate the entropy associated with
black holes in the $f(R)$ theory that have a constant Ricci scalar, one can obatin \cite{Cognola:2005de}
\begin{equation} \label{ent}
S=\frac{1}{4}Af_R(R)\mid_{r=r_h},\end{equation}
where A is the area of  the event horizon. Using Eqs. (\ref{4d}) and (\ref{nd}), we obtain the entropy as
\begin{equation} \label{s4}
  {S=\pi r_{h}{}^2(1-16b\Lambda) ,\quad{\textrm when}\;  N=4,  \qquad S=\frac{\Omega_{N-2} r_{h}{}^{N-2}(1-16b\Lambda_{eff})}{4\pi},  \quad {\textrm when}\; N>4} \end{equation}
where $\Omega_{N-2}$  denotes the volume of the unit (N-2)-sphere. {We plot the behavior of  entropy in Figures \ref{Fig:2} \subref{fig:2b} and \subref{fig:2e}.  It should be stressed that for positive entropy especially for positive cosmological constant, the dimensional parameter $b$ must be  $b<\frac{1}{16\Lambda}$ when $N=4$ and $b<\frac{1}{16\Lambda_{eff}}$ for $N>4$. This puts a constraint on the parameter $b$ \cite{Nunes:2016drj}.}

Finally, it is  known that there are several ways to study the stability of a black hole \cite{Nashed:2003ee}: among these approaches is the thermodynamic stability which is related to the sign of its heat capacity $C_h$. Now, we are going to analyze the thermal stability of black hole solutions via the behavior of their heat capacities \cite{Nouicer:2007pu,2011GrCo...17...35D,Chamblin:1999tk}
\begin{equation}\label{m55}
C_h= \frac{\partial M_h}{\partial r_h} \left(\frac{\partial T}{\partial r_h}\right)^{-1}.
\end{equation}
If the heat capacity $C_{h} > 0$ ($C_h < 0$), then the black hole is thermodynamically stable (unstable). Thus, a black hole  with a negative heat capacity is thermally unstable.
Using Eqs.  (\ref{tem}) and (\ref{m33}) in (\ref{m55}), we obtain
\begin{equation}\label{m555}
C_h= 4\pi r_h{}^2, \qquad {\textrm for}\; N=4; \qquad C_h= 4\pi r_h{}^{N-2}, \quad  {\textrm for} \; N>4.
\end{equation}

{The behavior of heat capacity is plotted in Figures \ref{Fig:2} \subref{fig:2b} and \subref{fig:2e}, which show that the black hole solutions (\ref{4d}) and (\ref{nd}) are stable when $\Lambda<0$. Of note the case of negative temperature has been discussed in \cite{Saridakis:2009uu,Nashed:2020kdb}.}

\section{Summary and discussion}\label{S6}

{Recently, without introducing any exotic matter, extended theory of gravity, $F(R)$, has been considered
as an alternative approach to explain the galactic rotation curves and the cosmic acceleration \cite{Nojiri:2010wj,Nojiri:2017ncd,DeFelice:2010aj}. The approach
results from effective theory aimed to deal with quantum fields in curved space-time at ultraviolet scales which give
rise to additional contributions with respect to GR also at infrared scales: in this perspective,
galactic, extra-galactic and cosmological scales can be affected by these gravitational corrections without requiring
large amounts of unknown material dark components. In the framework of $f(R)$, one may consider that the gravitational interaction acts differently at different scales, while the results of GR at Solar System scales are preserved. In other words, GR is a particular case of a more extended class of $f(R)$ gravitational theory. From a conceptual viewpoint, there is no
a priori reason to restrict the gravitational Lagrangian to a linear function of the Ricci scalar minimally coupled to
matter.}

{In this study, we  derived $N$-dimension, $N>4$, black hole solutions in $f(R)$ gravitation theory.}  We applied a spacetime, (which possesses a $k$-dimension Euclidean metric, $\ell$-dimension angular coordinates, and one unknown function of the radial coordinate) to the gravitational field equations in the $f(R)$ gravity with the quadratic form $f(R)=R+b R^2$. The resulting differential equations are solved exactly  without any assumption and general solutions for $N=4$ and $N>4$. These solutions are classified  as follows:\vspace{0.1cm}\\
i) The solution of $N=4$ is completely identical to GR and gives a planar black hole spacetime which is a singular one. \vspace{0.1cm}\\
ii) The solutions with $N>4$ are affected  by the higher curvature order, i.e., the solutions contain the dimensional parameter $b$.  Generally, the solutions in the case of $N>4$ cannot reduce to the GR solutions because the parameter $b$ is not allowed to go  to 0. Of note,  metric (\ref{m2}) satisfies $g_{tt}g_{rr}=1$, which, of course, is not the general one. We would like to emphasize that even if we use two different unknown functions, the solutions of the resulting differential equation will give $g_{tt}g_{rr}=1$ after some re-scaling.

To construct rotating black {hole} solutions, we  applied a coordinate transformation that relates the temporal coordinate  and the rotating one in the case of $N=4$ and between the temporal and  angular-coordinates in the case of $N>4$ and  derived the solutions of  rotating black {hole} that satisfy   gravitational field equations for $f(R)=R+bR^2$.  The topology of the output solution in the case of $N=4$ is a  cylindrical spacetime with $R\times S^1$ and  $0\leq \phi_{1}< 2\pi$ and $-\infty < z_1 < \infty$;  in the case of $N>4$ it is   $0\leq \phi_{\ell}< 2\pi$ and $-\infty < z_k < \infty$ in the case of $N>4$.

We studied the physics of the rotating black {hole} solutions by calculating their conserved quantities    using  Komar \cite{PhysRev.127.1411}.  This method provides divergent quantities of energy and angular momentum for  the two cases of $N=4$ and $N>4$.Thus, we used the regularization method to obtain
the energy-momentum and angular one with their finite values.
The regularization method used in this study is  relocalization, which is created from the change of the Lagrangian for the gravitational field in terms of the total derivative. From the method, the representations for
the energy-momentum and angular one were acquired. It was also  confirmed that these results were consistent with
those derived in \cite{Awad:2002cz,PhysRevD.86.024013}.

 Finally,  we  derived the entropies of black hole solutions (\ref{4d}) and (\ref{nd}) and showed that they were not proportional to the area of  horizons because of the existence of the $f_R$ term that is not trivial in our study \cite{PhysRevD.86.024013}. From these calculations, we put a constraint on the dimensional parameter $b$ to get positive entropy. Additionally, we  studied   heat capacity for  the cases of $N=4$ and $N>4$. We  showed   that the system is thermally stable, for $\Lambda>0$ and $\Lambda_{eff}>0$, in  both cases as shown in Figure \ref{fig:2c} \ref{fig:2c}. { Here we emphasize that the case of $N=4$  completely coincides with GR; however, if one considers the case $f(R)=R-2\alpha\sqrt{R}$ the situation is differerent \cite{Nashed:2020kdb} and one obtain a new solution in the case of $N=4$. A detailed analysis of this solution regarding thermodynamic analysis was performed \cite{Nashed:2020kdb} .} \vspace{0.3cm}

\vspace{0.3cm}

  {\centerline{\bf Appendix I:  Symbols used in the calculations of conserved quantities}}\vspace{0.3cm}

Indices ${ k, l, \cdots }$ are the coframe indices and  $\gamma$, $\delta$,
$\cdots$ are the  coordinate ones. The wedge product is represented by $\wedge$ and the
interior product   is given
by $\xi \rfloor \Psi$.  The coframe ${\vartheta}^{i}$ is defined as ${ \vartheta}^{i}={e^i}_\mu
dx^\mu$ and the frame $e_i$ is defined as ${ e_i={e_i}^\mu \partial_\mu}$
with ${ {e^i}_\mu}$ and ${ {e_i}^\mu} $ being the covariant of the vielbein and its
 contravariant, respectively. The volume is expressed by $\eta:=\vartheta^{\hat{0}}\wedge \vartheta^{\hat{1}}\wedge
\vartheta^{\hat{2}}\wedge\vartheta^{\hat{3}}$. In addition, we describe
\[{ \eta}_i:=e_i \rfloor \eta = \ \frac{1}{3!} \
\epsilon_{i j k l} \ { \vartheta}^j \wedge
{ \vartheta}^k \wedge { \vartheta}^l,\]
where $\epsilon_{ i j k l}$ is completely antisymmetric.

\vspace{0.3cm}

 {\centerline{\bf Appendix II:  Non-zero components for the Christoffel symbols of the second kind}} {\centerline{\bf and Ricci curvature tensor}}\vspace{0.3cm}

Using  Eq. (\ref{m2}), we find the non-zero components for the Christoffel symbols of the second kind and Ricci curvature tensor
  \begin{eqnarray*}
&& \Gamma^t{}_{t\; t}=-\Gamma^r{}_{r\; r}=\frac{h'}{2h},\qquad \qquad \Gamma^t{}_{r\; t}=\frac{hh'}{2}, \nonumber\\
 && \Gamma^{r}{}_{\phi_1\; {\phi_1}}=\Gamma^{r}{}_{\phi_2\; {\phi_2}}\cdots \cdots \Gamma^{r}{}_{\phi_{N-\ell}\; {\phi_{N-\ell}}}=\Gamma^{r}{}_{z_1\; {z_1}}=\Gamma^{r}{}_{z_2\; {z_2}}\cdots \cdots \Gamma^{r}{}_{z_{N-\ell-2}\; {z_{N-\ell-2}}}=-r h,\nonumber\\
&&\Gamma^{\phi_1}{}_{r\; {\phi_1}}=\Gamma^{\phi_2}{}_{r\; {\phi_2}}\cdots \cdots \Gamma^{\phi_{N-\ell}}{}_{r\; {\phi_{N-\ell}}}=\Gamma^{z_1}{}_{r\; {z_1}}=\Gamma^{z_2}{}_{r\; {z_2}}\cdots \cdots \Gamma^{z_{N-\ell-2}}{}_{r\; {\phi_{N-\ell-2}}}=\frac{1}{r}.
  \end{eqnarray*}
 \begin{eqnarray*}
 && R_{t\;r\;t\;r}=\frac{h''}{2}, \qquad R_{t\;\phi_1\;t\;\phi_1}= R_{t\;\phi_2\;t\;\phi_2}=\cdots  \cdots R_{t\;\phi_{N-\ell}\;t\;\phi_{N-\ell}}=R_{t\;z_1\;t\;z_1}= R_{t\;z_2\;t\;z_2}=\cdots  \cdots R_{t\;z_{N-\ell-2}\;t\;z_{N-\ell-2}}=\frac{r\;h h'}{2}, \nonumber\\
&& R_{r\;\phi_1\;r\;\phi_1}= R_{r\;\phi_2\;r\;\phi_2}=\cdots  \cdots R_{r\;\phi_{N-\ell-2}\;r\;\phi_{N-\ell-2}}=R_{r\;z_1\;r\;z_1}= R_{r\;z_2\;r\;z_2}=\cdots  \cdots R_{r\;z_{N-2}\;r\;z_{N-2}}=-\frac{r h'}{2h},\nonumber\\
&& R_{\phi_1\;\phi_2\;\phi_1\;\phi_2}= R_{\phi_1\;\phi_3\;\phi_1\;\phi_3}=\cdots \cdots R_{\phi_1\;\phi_{N-\ell}\;\phi_1\;\phi_{N-\ell}}= R_{\phi_2\;\phi_3\;\phi_2\;\phi_3}=R_{\phi_2\;\phi_4\;\phi_2\;\phi_4}\cdots \cdots=  \nonumber\\
 && R_{\phi_2\;\phi_{N-\ell}\;\phi_2\;\phi_{N-\ell}} \cdots \cdots R_{\phi_{N-\ell-1}\;\phi_{N-\ell}\;\phi_{N-\ell-1}\;\phi_{N-\ell}}= R_{z_1\;z_2\;z_1\;z_2}= R_{z_1\;z_3\;z_1\;z_3}=\cdots \cdots R_{z_1\;z_{N-\ell-2}\;z_1\;z_{N-\ell-2}}=\nonumber\\
 && R_{z_2\;z_{N-\ell-2}\;z_2\;z_{N-\ell-2}} \cdots \cdots R_{z_{N-\ell-3}\;z_{N-\ell-2}\;z_{N-\ell-3}\;z_{N-\ell-2}}=-r^2 h,\end{eqnarray*}
\newpage
\subsection*{Acknowledgments}
This work was supported in part by the Egyptian Ministry of Scientific Research under project No. 24-2-12. Moreover, the work of KB was partially supported by the JSPS KAKENHI Grant Number JP 25800136 and Competitive Research Funds for Fukushima University Faculty (18RI009).

%

\end{document}